\documentclass[english,aps,prl,twocolumn,graphics,graphicx,superscriptaddress]{revtex4-1}
\usepackage[latin9]{inputenc}
\usepackage{amsmath,mathtools}
\usepackage{amssymb}
\usepackage{graphicx}
\usepackage{verbatim}
\usepackage{soul}
\usepackage{bm}% bold math
\usepackage{gensymb}
\usepackage{diagbox}
\usepackage{siunitx}
\usepackage[T1]{fontenc}
\usepackage[english]{babel}
\usepackage{mathtools}
\usepackage{newtxtext}  % scaled=1.005
\usepackage{amsmath}
\usepackage{amsfonts}
\usepackage[varbb,varg]{newtxmath} % scaled=1.005
\usepackage{bm,dsfont,eucal}
\newcommand{\bk}{\boldsymbol{k}}

\usepackage{physics,braket,siunitx,slashed}
\usepackage[colorlinks=true,
            linkcolor=blue,   % equations, figures, sections, etc.
            citecolor=blue,   % bibliography citations
            urlcolor=blue,    % URLs
            breaklinks=true]{hyperref}

\renewcommand\vec{\boldsymbol}

\begin{document}
\title{Orbital Magnetization and Streda Formula of Interacting Electrons in the Mean-field Approximation}

\author{Jihang Zhu}
\affiliation{Department of Materials Science and Engineering, University of Washington, Seattle, Washington 98195, USA}
\author{Chunli Huang}
\affiliation{Department of Physics and Astronomy, University of Kentucky, Lexington, Kentucky 40506-0055, USA}
%\date{October 2023}
\date{\today} 

\begin{abstract}
We study the magnetic-field response of interacting electron systems within mean-field theory using perturbation theory. We show that the linear response of the mean-field density-matrix to a weak magnetic field is purely geometric: it depends only on wavefunction derivatives, the Berry connections linking the occupied and unoccupied subspaces, and does not explicitly depend on the interaction potential and the quasiparticle dispersion.  
This leads to compact, gauge-invariant projector expressions for both the St\v{r}eda formula and the formula for orbital magnetization.  
Our calculation explicitly elucidates the role of exchange and self-consistency in defining current vertices for orbital magnetization calculations. Our work establishes a direct connection between mean-field theory, quantum geometry and the non-interacting topological band theory.
\end{abstract}

\maketitle

\section{Introduction}

Topological band theory \cite{DXiao_RMP_2010, MZHasan_TI_2010, Vanderbilt_2018, ABansil_TI_2016} extends the successful Bloch band theory by showing that certain physical properties of electronic systems are governed not by local details of the band structure but by the global topology of the Bloch bands.   The most prominent example is the Thouless-Kohmoto-Nightingale-den Nijs (TKNN) formula \cite{TKNN_1982}, which expresses the transverse Hall conductivity as a Brillouin-zone integral of the Berry curvature associated with the occupied Bloch states.  
Because these integrals yield quantized topological invariants (Chern number) \cite{Kohmoto_1985} the corresponding physical observables are insensitive to small perturbations that do not close the energy gap, leading to the robust quantization of the Hall resistance. Beyond topology, recent research has increasingly focused on the \emph{local} structure of Bloch states in momentum space, which encodes rich geometric information through the Berry curvature and the quantum metric~\cite{Provost_1980, Resta_2011, onishi2024fundamental, shinada2025quantum, Slager_2023, gao2025quantum, avdoshkin2025multistate, mitscherling2025gauge}.
The emerging framework of quantum geometry~\cite{gao2025quantum} extends topological band theory by emphasizing that, beyond topological quantization, a wide range of physical properties are determined by the geometric structure of Bloch wave functions in momentum space.

In most discussions, topological band theory is developed for non-interacting electrons, where the Berry curvature and related geometric quantities can be computed directly from single-particle Bloch wave functions.   
In this work, we show that several central results of topological band theory, including the St\v{r}eda formula~\cite{Streda_1982} and the modern theory of orbital magnetization~\cite{JShi_OM_2007, Thonhauser_OM_2005, Xiao_OM_2005, Ceresoli_OM_2006, raoux2015orbital, chen2011unified, gonze2011density}, can be naturally generalized to mean-field quasiparticles by replacing the non-interacting Bloch energies and wave functions with their self-consistent mean-field counterparts.  
The mean-field case is more subtle because the quasiparticle states depend self-consistently on the properties of all other occupied states. 

Our key result is that the linear response of the mean-field density matrix to a weak magnetic field depends only on the interband Berry connections between occupied and unoccupied manifolds, and not explicitly on the interaction potential or the detailed band dispersion. This suggests a quantum-geometric origin and it remains robust in the presence of interactions treated at the mean-field level. Our work thus provides a unified framework connecting mean-field quasiparticle with topological band theories for non-interacting systems.  
It also offers theoretical justification for prior Hartree-Fock studies of rhombohedral multilayer graphene~\cite{das2025momentum, das2025microscopic, das2024unconventional}, where orbital magnetization was computed directly from the mean-field band structure.

Prior to our work, Kang.~et.~al~\cite{kang2025orbital} recently derived an expression for the orbital magnetization in Hartree-Fock theory using Green's function approach \cite{Nourafkan_OM_2014}, based on an expansion of the magnetic-flux in coordinate space similar to Ref.~\cite{raoux2015orbital, chen2011unified, gonze2011density}. Our derivation is based on the more straightforward expansion of the vector potential in the momentum space, similar to Ref.~\cite{JShi_OM_2007}, and highlights the geometric origin of the response in terms of mean-field projectors.

\section{Many-body Perturbation theory in mean-field approximations}
We first briefly review the perturbation theory of the Hartree-Fock equations.
Consider a many-electron system described at the mean-field level, characterized by a one-body density matrix (which is equivalently a projector onto the occupied states),
\begin{equation}
    P = \sum_i \ketbra{i}{i}.
\end{equation}
Throughout this section, we use the following notation:
\begin{equation}
    i,j:~\text{occupied states}, \qquad 
    m,n:~\text{unoccupied states}.
\end{equation}
Because $P^2 = P$, and since the final expressions depend only on the Berry connection, we will refer to $P$ as the \emph{projector} rather than the density matrix. The mean-field (MF) Hamiltonian is defined as
\begin{equation}
    H^{\mathrm{MF}} \equiv H^{\mathrm{MF}}[P] = T + \Sigma[P],
\end{equation}
where $T$ denotes the kinetic energy operator and $\Sigma[P]$ is the mean-field self-energy, which is a functional of $P$. The stationary mean-field condition requires that $P$ commute with $H^{\mathrm{MF}}$,
\begin{equation} \label{eq:stationary_cond}
    [H^{\mathrm{MF}}, P] = 0.
\end{equation}

We now perturb the system by a weak external potential $H'$, under which
\begin{align}
    P \to P + \delta P, \qquad 
H^{\mathrm{MF}}[P] \to H^{\mathrm{MF}}[P + \delta P] + H'.
\end{align}
Because $P$ is a projector, linearizing the constraint $P^2 = P$ (and equivalently $(P+\delta P)^2 = P + \delta P$) implies that $\delta P$ is purely off-diagonal between the occupied and unoccupied sectors:
\begin{align}
    P\,\delta P\,P &= 0, \qquad (1-P)\,\delta P\,(1-P) = 0.
\end{align}
Hence,
\begin{align}
    \delta P &= \sum_{i}\!\left(\ketbra{\delta i}{i} + \ketbra{i}{\delta i}\right) \nonumber\\
    &= \sum_{m i} \!\left( X_{mi}\ketbra{m}{i} + X_{mi}^*\ketbra{i}{m} \right),
\end{align}
where the perturbed occupied states are expanded in the unoccupied basis $\ket{\delta i} = \sum_m X_{mi}\ket{m}$ and the coefficients $X_{mi} = \braket{m|\delta i}$ are the unknown complex amplitudes.

The coefficients $X_{mi}$ can be obtained by linearizing the stationary condition, Eq.~\eqref{eq:stationary_cond}, which yields
\begin{equation}
    [\,T + \Sigma[P],\, \delta P\,] = -[\,\Sigma[\delta P] + H',\, P\,].
\end{equation}
This equation describes the new self-consistent condition in the presence of the perturbation $H'$, accurate to the first order in $H'$. 
Here we have used the linearized dependence of the Hartree-Fock self-energy, $\Sigma[P + \delta P] = \Sigma[P] + \Sigma[\delta P]$. Taking matrix elements between unoccupied ($m$) and occupied ($i$) states yields
\begin{equation}
    (\varepsilon_m - \varepsilon_i)\,X_{mi}
    + (n_i - n_m)\,\Sigma_{mi}[\delta P]
    = - (n_i - n_m)\,H'_{mi},
\end{equation}
where $\varepsilon_{m/i}$ denote the mean-field eigenvalues and $n_{m/i}$ are the corresponding occupation numbers at the stationary point.

In the ``frozen self-energy'' approximation, $\Sigma_{mi}[\delta P]\to 0$, this reproduces the familiar first-order result of the single-particle perturbation theory,
$X_{mi} = -H'_{mi}/(\varepsilon_m-\varepsilon_i)$. 
However, once the self-energy is allowed to vary, a single particle--hole excitation $i \!\rightarrow\! m$ can kick out all other occupied states $j$, and put them in some unoccupied states $n$ that have the same conserved quantum numbers.  
Including this induced variation of the self-energy, one obtains
\begin{align} \label{eq:stability_cond}
X_{mi}\!-\!
    \chi^0_{mi}\!\sum_{nj}\!\left(
    \!\braket{mj|\bar V|in}\!X_{nj}
    \!+\!\braket{mn|\bar V|ij}\!X_{nj}^*
    \!\right) \!
    =\!\chi^0_{mi} H_{mi}',
\end{align}
where $\chi^0_{mi} = (n_m - n_i)/(\varepsilon_m - \varepsilon_i)$ is the particle-hole susceptibility defined in terms of Hartree-Fock energies. The notation for the matrix elements, $\braket{mj|\bar V|in}=\braket{mj|V|in}-\braket{mj|V|ni}$, contains both direct and exchange scatterings. The linear system above has dimension $2N_{\mathrm{emp}}N_{\mathrm{occ}}$, where the factor of $2$ accounts for the real and imaginary components of the complex amplitudes $X_{mi}$.  
We note that the Hartree-Fock solution by construction satisfies the stationary condition, Eq.~\eqref{eq:stationary_cond}, but is not necessarily stable.  
The stability is determined by the homogeneous part ($H'_{mi}=0$) of Eq.~\eqref{eq:stability_cond}.  
If the corresponding eigenvalue spectrum is positive definite, the Hartree-Fock solution is stable.  
This requirement is known as the Thouless stability criterion in Hartree-Fock theory~\cite{thouless2013quantum}. It is therefore not surprising that the same equation naturally emerges when the Hartree-Fock solution is perturbed with $H'$.

\section{Mean-Field Projector Induced by a Magnetic Field}\label{sec:projector}
We now focus on the cell-periodic part of the Bloch functions, such that the mean-field projector is
\begin{equation}
    P = \sum_{v\vec k} \ketbra{u_{v\vec k}}{u_{v\vec k}},
\end{equation}
The mean-field Hamiltonian 
$H^{\mathrm{MF}}_{\vec k}\ket{u_{n\vec k}} = \varepsilon_{n\vec k}\ket{u_{n\vec k}}$, where the band index $n$ labels the mean-field eigenstates, with $n = v$ and $n = c$ denoting, respectively, the occupied and unoccupied states. 
Then we consider a weak external vector potential with Fourier component $\vec{A}_{\vec q}$, which couples to the electrons through the bare velocity operator,
\begin{equation}
    \hat{\vec v}_{\vec k} = \frac{1}{\hbar}\frac{\partial T}{\partial \vec k}.
\end{equation}
The corresponding matrix elements are
\begin{equation}
    H'_{c\vec k+\vec q,\, v\vec k} 
    = e\,\vec{A}_{\vec q}\!\cdot\!\vec v_{c\vec k+\vec q,\, v\vec k},
\end{equation}
where $\vec v_{c\vec k+\vec q,\, v\vec k}
=(\bra{u_{c\vec k+\vec q}}\hat{\vec v}_{\vec k+\vec q}\ket{u_{v\vec k+\vec q}} + \bra{u_{c \vec k}}\hat{\vec v}_{\vec k}  
\ket{u_{v\vec k}})/2$ 
is the \emph{bare current vertex} (or photon vertex), which describes the transition of an occupied state $\ket{u_{v\vec k}}$ to an empty state $\ket{u_{c,\vec k+\vec q}}$ induced by the absorption of a photon with momentum $\vec q$.

The velocity operator is defined as the derivative of the kinetic part of the Hamiltonian because interaction terms that are local in real space, like the Coulomb potential, do not couple to vector-potential via the minimal coupling. 
The perturbation $H'$ leads to the following change in density-matrix:
\begin{equation}\label{eq:delta_P}
\begin{split}
\delta P_{\vec q} = \sum_{cv\vec k} X_{c\vec k+\vec q,v\vec k} \ketbra{u_{c\vec k+\vec q}}{u_{v\vec k}} + X_{c\vec k-\vec q,v\vec k}^* \ketbra{u_{v\vec k}}{u_{c\vec k-\vec q}}.
\end{split}
\end{equation}
Note the density matrix $\delta P_{\vec q} = \delta P_{-\vec q}^\dagger$, so that the real-space density matrix is Hermitian.  This form of $\delta P_{\vec q}$ also corresponds to the creation operator of a bosonic excitation with momentum $\vec q$ in the equation-of-motion (or time-dependent mean-field) theory. \cite{wolf2024quasi}. The backward amplitudes  $X^{*}_{c,\vec k-\vec q;,v,\vec k} $ are normally defined as $Y_{c,\vec k-\vec q;,v,\vec k}$, and the pair $(X, Y)$ together constitute a consistent eigenmode of the linearized dynamics.

It is convenient to rewrite the inhomogeneous stability equation [Eq.~\eqref{eq:stability_cond}] as a self-consistent vertex equation by parameterizing the response amplitude as
\begin{equation}
\label{eq:X_and_Gamma}
    X_{c\vec k+\vec q,\, v\vec k}
    = \chi^0_{c\vec k+\vec q,\, v\vec k}\, e\,\vec{\Gamma}_{c\vec k+\vec q,\, v\vec k}\!\cdot\!\vec{A}_{\vec q},
\end{equation}
where $\chi^0_{c\vec k+\vec q,\, v\vec k} = (n_{c\vec k+\vec q} - n_{v\vec k})/(\varepsilon_{c\vec k+\vec q} - \varepsilon_{v\vec k})$ is the susceptibility defined in terms of mean-field quasiparticles.  
Substituting Eq.~\eqref{eq:X_and_Gamma} into Eq.~\eqref{eq:stability_cond} gives the self-consistent vertex equation,
\begin{align}
&\vec{\Gamma}_{c\vec k+\vec q,\, v\vec k}
    = \vec v_{c\vec k+\vec q,\, v\vec k} + \sum_{c' v' \vec{k}'} \chi^0_{c'\vec k'+\vec q,\, v'\vec k'}\times \nonumber \\
    &\Big[
        \braket{u_{c,\vec k+\vec q},\, u_{v'\vec k'}|\bar{V}|u_{v\vec k}, u_{c',\vec k'+\vec q}}
        \vec{\Gamma}_{c'\vec k'+\vec q,\, v'\vec k'}  \nonumber\\
    &
        + \braket{u_{c,\vec k+\vec q},\, u_{c',\vec k'-\vec q}| \bar{V} | 
         u_{v\vec k}, u_{v'\vec k'}}
        \vec{\Gamma}^*_{c'\vec k'-\vec q,\, v'\vec k'}
    \Big].
\end{align}
This expression has a natural diagrammatic interpretation: it represents the static limit of the Bethe-Salpeter equation for the current vertex dressed by particle-hole interactions.  
The first term on the right-hand side is the bare vertex, while the second term captures vertex corrections resulting from the repeated direct and scattering of particle-hole excitations. 

The key step in our derivation is that we find a closed-form solution to the vertex equation in the long-wavelength limit ($q \to 0$) [see Appendix]:
\begin{equation} \label{eq:vertex_q=0}
    \vec{\Gamma}_{c\vec k,\, v\vec k}
    = \frac{1}{\hbar}\,
      \braket{u_{c\vec k}|\nabla_{\!\vec k} H^{\mathrm{MF}}|u_{v\vec k}}.
\end{equation}
Furthermore, because the momentum derivative of the mean-field Hamiltonian is related  to the interband Berry connection,
\begin{equation}
    -\frac{ 
        \braket{u_{c\vec k}|\nabla_{\!\vec k} H^{\mathrm{MF}}|u_{v\vec k}}
    }{
        \varepsilon_{c\vec k} - \varepsilon_{v\vec k}
    }
    = \braket{u_{c\vec k}|\nabla_{\!\vec k} u_{v\vec k}},
\end{equation}
the response matrix elements take a remarkably simple form,
\begin{equation}\label{eq:X_simple}
    X_{c\vec k,\, v\vec k}
    = \frac{e}{\hbar}\braket{u_{c\vec k}|\nabla_{\!\vec k} u_{v\vec k}}\!\cdot\!\vec{A}_{\vec q}.
\end{equation}
Notably, this result contains no explicit dependence on either the interaction potential $V$ or the quasiparticle energies $\varepsilon_{n\vec k}$. It is entirely determined by the interband Berry connection between the occupied and unoccupied states. Equation~\eqref{eq:X_simple} is not gauge invariant in both the momentum space ($\ket{u_{v\vec k}}\rightarrow e^{i\theta_k}\ket{u_{v\vec k}}$) and the coordinate space ($\vec{A}_q\rightarrow\vec{A}_q+\vec q\chi$).

Substituting this solution back to the perturbed density matrix Eq.~(\ref{eq:delta_P}), we obtain

\begin{align} \label{eq:delta_Pq}
    \delta P_{\vec q }
    = \frac{e}{\hbar}\sum_{cv\vec k} 
    \Big(&\vec{A}_{\vec q}\!\cdot\!\braket{u_{c\vec k}|\nabla_{\!\vec k} u_{v\vec k}}
        \ket{u_{c,\vec k+\vec q}}\bra{u_{v\vec k}}  \nonumber\\
    &+ \vec{A}^*_{-\vec q} \cdot
    \braket{u_{c\vec k}|\nabla_{\!\vec k} u_{v\vec k}}^* 
    \ket{ u_{v \vec k}} \bra{u_{c\vec k-\vec q}} \Big).
\end{align}

This expression is remarkable in that it does not depend explicitly on the microscopic details of the band dispersion or the interaction. 
The magnetic field induced change in the mean field projector is given entirely by the Berry connections and their momentum-space derivatives of the mean-field Bloch states.  
We would obtain an identical expression by applying perturbation theory to the response of non-interacting electrons to an external magnetic field~\cite{JShi_OM_2007}, with the mean-field wavefunctions mapping back to the non-interacting wavefunctions $\ket{u_{n\vec k}}\!\rightarrow\!\ket{u_{n\vec k}^{0}}$.  Fundamentally, this simplification arises because Eq.~\eqref{eq:vertex_q=0} is simply 
a solution of the Ward identity when the Hartree-Fock approximation is used for the Green's function,
\begin{equation}
    \hbar \vec q \cdot \vec{\Gamma}_{\vec k+\vec q,\vec k}
    =  G^{-1}_{\vec k}- G^{-1}_{\vec k+\vec q}
    = \vec q \cdot \nabla_{\!\vec k} H^{\mathrm{MF}}_{\vec k} + O(q^2).
\end{equation}
This follows from the fact that the Hartree-Fock approximation is a conserving approximation \cite{baym1961conservation}.
Since the Ward identity originates from the continuity equation, this connection suggests that the underlying conservation laws impose nontrivial constraints on how many-body interactions modify the quantum geometry in momentum space. We now use Eq.~\eqref{eq:delta_Pq} to derive the 
St\v{r}eda formula and orbital magnetization.

\subsection{St\v{r}eda formula}

The St\v{r}eda formula states that the change of electron density at fixed chemical potential is a topological quantity,
\begin{equation}
    \frac{dn_e}{dB}\bigg|_{\mu}=\frac{C}{\phi_0},
\end{equation}
where $C$ is the total Chern number of the occupied bands and $\phi_0 = h/e$ is the flux quantum. 

To derive this relation, we compute the change of the density matrix $\delta P_{\vec q}$ to linear order in the magnetic field. Expanding $\Tr(\delta P_{\vec q})$ to linear order in $\vec q$, we use
\begin{align}
\braket{u_{v\vec k}|u_{c,\vec k+\vec q}} 
&= q_i \braket{u_{v\vec k}|\partial_{k_i} u_{c\vec k}} + \mathcal{O}(q^2), \\
\braket{u_{c,\vec k-\vec q}|u_{v\vec k}} 
&= q_i \braket{u_{c\vec k}|\partial_{k_i} u_{v\vec k}} + \mathcal{O}(q^2),
\end{align}
together with $\partial_{k_i}\braket{u_{c\vec k}|u_{v\vec k}}=0$. Substituting into Eq.~\eqref{eq:delta_Pq}, we obtain
\begin{align}
\Tr \delta P_{\vec q}
=
\frac{2ie}{\hbar} q_i A_j(\vec q)
\sum_{cv\vec k}
\Im\!\left[
\braket{u_{c\vec k}|\partial_{k_j} u_{v\vec k}}
\braket{u_{v\vec k}|\partial_{k_i} u_{c\vec k}}
\right].
\label{eq:deltaP_im}
\end{align}
Here we used $A_j(\vec q)=A_j^*(-\vec q)$ for a real vector potential. We now rewrite the sum over unoccupied states using the projectors
\begin{equation}
P_{\vec k}=\sum_v \ket{u_{v\vec k}}\bra{u_{v\vec k}},
\qquad
Q_{\vec k}=\sum_c \ket{u_{c\vec k}}\bra{u_{c\vec k}},
\end{equation}
with $P_{\vec k}+Q_{\vec k}=1$. Then,
\begin{align}
\Tr\delta P_{\vec q}
&=-\frac{2ie}{\hbar} q_i A_j(\vec q)
\sum_{v\vec k}
\Im\!\left[
\braket{\partial_{k_i} u_{v\vec k}|
Q_{\vec k}|\partial_{k_j} u_{v\vec k}}
\right].
\label{eq:deltaP_Q}
\end{align}

The matrix element
\begin{equation}
\chi^{(v)}_{ij}(\vec k)
=
\braket{\partial_{k_i} u_{v\vec k}|
Q_{\vec k}|\partial_{k_j} u_{v\vec k}}=
g^{(v)}_{ij}
-\frac{i}{2}\Omega^{(v)}_{ij}.
\end{equation}
is the quantum geometric tensor. Its real and imaginary parts correspond to the quantum metric and Berry curvature, respectively.
Substituting into Eq.~\eqref{eq:deltaP_Q}, we obtain
\begin{align}
\Tr\delta P_{\vec q}
=
\frac{ie}{\hbar} q_i A_j(\vec q)
\sum_{v\vec k}
\Omega^{(v)}_{ij}(\vec k).
\label{eq:deltaP_berry}
\end{align}

Using $B_l(\vec q)=i\epsilon_{l i j} q_i A_j(\vec q)$ and noting that only the antisymmetric part of $q_i A_j$ contributes upon contraction with $\Omega_{ij}$, we arrive at
\begin{align}
\Tr\delta P_{\vec q}
&=\frac{e}{2\hbar}
\epsilon_{l i j} B_l(\vec q) \sum_{v\vec k}\Omega^{(v)}_{ij}(\vec k) \\
&=
\frac{e}{\hbar}\mathbf{B}(\vec q)\cdot 
\sum_{v\vec k} \boldsymbol{\Omega}^{(v)}(\vec k),
\end{align}
where we introduced the Berry curvature vector
\begin{equation}
\Omega^{(v)}_l(\vec k)
=
\frac{1}{2}\epsilon_{l i j}\,\Omega^{(v)}_{ij}(\vec k).
\end{equation}
In two dimensions, both $\mathbf{B}$ and $\boldsymbol{\Omega}$ point along the $z$ direction (perpendicular to the electron motion plane). We therefore obtain the St\v{r}eda relation,
\begin{equation}
\frac{dn_e}{dB}\bigg|_{\mu}
=
\frac{\Tr\delta P_{\vec q}}{\text{Area}\times B_z(\vec q)}
=
\frac{eC}{h},
\end{equation}
\begin{equation}
C = \frac{1}{2\pi}\int_{\mathrm{BZ}} \sum_v \Omega^{(v)}_{z}(\vec k)
d^2\vec k.
\end{equation}
$C$ is the Chern number arising from the total Berry curvature of the occupied states.

%For a two-dimensional system, only the out-of-plane component contributes, and Eq.~\eqref{eq:deltaP_3D} reduces to
%\begin{equation}
%\Tr\delta P_{\vec q}
%=
%\frac{e}{\hbar}
%B_z(\vec q)
%\sum_{v\vec k}
%\Omega^{(v)}_{z}(\vec k),
%\end{equation}
%where $\Omega^{(v)}_{z} \equiv \Omega^{(v)}_{xy}$.

\subsection{Orbital Magnetization}
In mean-field theory, the total energy is given by
\begin{equation}
E = \Tr\!\left[\big(T + \tfrac{1}{2}\Sigma[P]\big)P\right],
\end{equation}
where $T$ is the one-body kinetic energy operator and $\Sigma[P]$ is the self-energy functional of the projector $P$.  When an external perturbation $H'$ is applied, the mean-field projector changes as $P \rightarrow P + \delta P$.  
Expanding the total energy to the first order in $H'$, we obtain
\begin{align}
\delta E
&= \Tr\!\left[\big(T + \tfrac{1}{2}\Sigma[P]\big)\delta P\right]
 + \Tr\!\left[\tfrac{1}{2}\Sigma[\delta P]\,P\right]
 + \Tr\!\left[H' P\right] \nonumber\\
&= \Tr\!\left[\big(T + \Sigma[P]\big)\delta P\right]
   + \Tr\!\left[H' P\right] \nonumber\\
&= \Tr\!\left[H^{\mathrm{MF}}[P]\,\delta P\right]
   + \Tr\!\left[H' P\right],
\end{align}
where we used $\Tr\!\left[\Sigma[\delta P]P\right]=\Tr\!\left[\Sigma[ P]\delta P\right]$. Here
$H^{\mathrm{MF}}[P] = T + \Sigma[P]$ is the stationary mean-field Hamiltonian in the absence of perturbation.

Introducing the grand-canonical operator
\begin{equation}
\hat{\Omega}=H^{\mathrm{MF}}[P]-\mu N,
\end{equation}
we evaluate ${\rm Tr}(\hat{\Omega}\,\delta P_{\vec q})$ to leading order in the magnetic field. Using
\begin{align}
\braket{u_{v\vec k}|\Omega|u_{c,\vec k+\vec q}} 
&= q_i \braket{u_{v\vec k}|\partial_{k_i} u_{c\vec k}}\frac{\xi_{c\vec k}+\xi_{v\vec k}}{2} + \mathcal{O}(q^2), \\
\braket{u_{c,\vec k-\vec q}|\Omega|u_{v\vec k}} 
&= q_i \braket{u_{c\vec k}|\partial_{k_i} u_{v\vec k}}\frac{\xi_{c\vec k}+\xi_{v\vec k}}{2}+ \mathcal{O}(q^2),
\end{align}
we find, 
\begin{align}
{\rm Tr}(\hat{\Omega}\,\delta P_{\vec q})
=-\frac{2ie}{\hbar} &q_i A_j(\vec q)
\sum_{cv\vec k}\bigg\{ \left(\frac{\xi_{c\vec k}+\xi_{v\vec k}}{2}\right)\times \nonumber \\
&\Im\!\left[
\braket{\partial_{k_i} u_{v\vec k}|
u_{c\vec k}\rangle\langle u_{c \vec k}|\partial_{k_j} u_{v\vec k}}
\right].\bigg\}
\end{align}
This expression closely resembles the St\v{r}eda formula in Eq.~\eqref{eq:deltaP_im}. However, due to the energy factor $(\xi_{c\vec k}+\xi_{v\vec k})/2$, it cannot in general be expressed purely in terms of the Berry curvature of the occupied states. Instead, it necessarily depends on matrix elements involving unoccupied states. As in the St\v{r}eda formula derivation, only the antisymmetric part of $q_i A_j(\vec q)$ contributes, since the quantity
\begin{align}
&\Im\!\left[
\braket{\partial_{k_i} u_{v\vec k}|
u_{c\vec k}\rangle\langle u_{c \vec k}|\partial_{k_j} u_{v\vec k}}
\right]\nonumber \\
&=-\Im\!\left[
\braket{\partial_{k_j} u_{v\vec k}|
u_{c\vec k}\rangle\langle u_{c \vec k}|\partial_{k_i} u_{v\vec k}}
\right]
\end{align}
is antisymmetric under $i \leftrightarrow j$. The real part is symmetric under $i \leftrightarrow j$. Using $iA_i(\vec q)q_j=\epsilon_{lij}B_l(\vec q)/2$,  we obtain
\begin{align}
{\rm Tr}(\hat{\Omega}\,\delta P_{\vec q})
=-\frac{e}{\hbar} &B_l(\vec q)\epsilon_{lij}
\sum_{cv\vec k}\bigg\{ \left(\frac{\xi_{c\vec k}+\xi_{v\vec k}}{2}\right)\times \nonumber \\
&\Im\!\left[
\braket{\partial_{k_i} u_{v\vec k}|
u_{c\vec k}\rangle\langle u_{c \vec k}|\partial_{k_j} u_{v\vec k}}
\right].\bigg\}
\end{align}

From thermodynamics, the total magnetization per unit area is defined as 
\begin{align}
    M_l =& -\frac{1}{\mathrm{Area}}\frac{\delta \Omega}{\partial B_l}\Big|_{B=0}=-\frac{1}{\mathrm{Area}}\frac{{\rm Tr}(\hat{\Omega}\,\delta P_{\vec q})}{ B_l(\vec q)}\\
    =&
    \frac{e}{\hbar} \epsilon_{lij}
\sum_{cv}\int\frac{d^2k}{4\pi^2}\left(\frac{\xi_{c\vec k}+\xi_{v\vec k}}{2}\right)\Im\!\left[
\braket{\partial_{k_i} u_{v\vec k}|
u_{c\vec k}\rangle\langle u_{c \vec k}|\partial_{k_j} u_{v\vec k}}
\right]
\end{align}
This expression gives the zero-temperature orbital magnetization and coincides with the well-known semiclassical result for non-interacting Bloch states
\cite{JShi_OM_2007, Thonhauser_OM_2005, Xiao_OM_2005, Ceresoli_OM_2006}. Comparing this expression with the St\v{r}eda formula, the orbital magnetization can be understood as an energy-weighted sum over occupied Bloch states executing infinitesimal closed loops in momentum space, with weight $(\xi_{v\vec k}+\xi_{c\vec k})/2$.
The Fermi-surface do not contribute to OM at zero temperature precisely because of this energy-weighting factor.
It is informative to contrast this result with the spin magnetization
\begin{equation}
    M_s = g\mu_B \langle \vec S \rangle,
\end{equation} 
where $\langle \vec S \rangle$ is the ground-state spin density and depends solely on properties of the occupied states. By contrast, orbital magnetization depends on both occupied and unoccupied states. This is because, fundamentally, it reflects the energy cost associated with mixing the occupied states with unoccupied states.

We show in the Appendix that this formula can also be written in the form that correspond to previous results \cite{JShi_OM_2007, Thonhauser_OM_2005, Xiao_OM_2005, Ceresoli_OM_2006}:
\begin{align}
    M_z = 
    \frac{e}{2\hbar} \int\frac{d^2k}{4\pi^2} 
    \sum_{v}
    \epsilon_{\alpha\beta}
    \Im \!
    \left[
    \bra{\partial_\alpha u_{\bk v}}
    \left( H_{\bk} + \varepsilon_{\bk v} - 2\mu \right)
    \ket{\partial_\beta u_{\bk v}}
    \right] .
\end{align}

\section{Conclusion}

We have demonstrated that, within mean-field theory, the change of the density matrix induced by minimal coupling to a weak vector potential $\vec{A}$ takes an elegantly simple form: it is given by the dot product between the inter-band Berry connection and the applied vector potential, $\delta P_{ck,vk}=\frac{e}{\hbar}\langle u_{ck}|\nabla u_{vk}\rangle\cdot \vec A$.
This structure is identical to that obtained for non-interacting electrons, and we have explicitly shown that it continues to hold in the presence of interactions at the mean-field level.  Our calculation emphasizes that, although the vector potential couples to the bare velocity vertex determined solely by the kinetic Hamiltonian the response of the density matrix is given by the fully dressed current vertex that incorporates the self-energies.
Building on this result, we derived expressions for both the orbital magnetization and the St\v{r}eda formula.

\textit{Note added:} During the completion of this manuscript, we became aware of a recent work by Ref.~\cite{XLiu_OMHF_2025} that discusses similar approach to derive orbital magnetization.

\textit{Acknowledgement:} We thank Oskar Vafek for sharing his manuscript \cite{kang2025orbital} with us, which inspired the present work.

\bibliographystyle{apsrev4-2}
\bibliography{references}

\newpage
\clearpage

\appendix

\section{Solution of the vertex equation for $\mathbf q=0$}

In the uniform limit (\(\vec q \to 0\)), the self-consistent vertex equation takes the form
\begin{align}\label{eq:Gamma_q0_appendix}
&\vec{\Gamma}_{c\vec k,\, v\vec k}
    = \vec v_{c\vec k,\, v\vec k} + \sum_{c' v' \vec{k}'} \chi^0_{c'\vec k',\, v'\vec k'}\times \nonumber \\
    &\Big[
        \braket{u_{c\vec k},\, u_{v'\vec k'}|\bar{V}|u_{v\vec k}, u_{c',\vec k'}}
        \vec{\Gamma}_{c'\vec k',\, v'\vec k'}  \nonumber\\
    &
        + \braket{u_{c\vec k},\, u_{c',\vec k'}| \bar{V} | 
         u_{v\vec k}, u_{v'\vec k'}}
        \vec{\Gamma}^*_{c'\vec k'\, v'\vec k'}
    \Big].
\end{align}

We now show that
\begin{equation}\label{eq:Gamma_exact_solution}
\vec{\Gamma}_{c\vec k,\, v\vec k}
=
\frac{1}{\hbar}
\braket{u_{c\vec k}|\nabla_{\!\vec k} H^{\mathrm{MF}}(\vec k)|u_{v\vec k}}
\end{equation}
is an exact solution of Eq.~\eqref{eq:Gamma_q0_appendix}. Using
\begin{align}
\chi^0_{c\vec k,\, v\vec k}\,\vec{\Gamma}_{c\vec k,\, v\vec k}
&=
-\frac{1}{\varepsilon_{c\vec k}-\varepsilon_{v\vec k}}
\frac{1}{\hbar}
\braket{u_{c\vec k}|\nabla_{\!\vec k}H^{\mathrm{MF}}(\vec k)|u_{v\vec k}}
\nonumber\\
&=
\frac{1}{\hbar}\braket{u_{c\vec k}|\nabla_{\!\vec k}u_{v\vec k}},
\label{eq:chiGamma_identity}
\end{align}
the interaction term on the right-hand side of Eq.~\eqref{eq:Gamma_q0_appendix} becomes
\begin{align}
\frac{1}{\hbar}\sum_{c'v'\vec k'}
\Big[
&\braket{u_{c\vec k},u_{v'\vec k'}|\bar V|u_{v\vec k},u_{c'\vec k'}}
\braket{u_{c'\vec k'}|\nabla_{\!\vec k'}u_{v'\vec k'}}
\nonumber\\
&+
\braket{u_{c\vec k},u_{c'\vec k'}|\bar V|u_{v\vec k},u_{v'\vec k'}}
\braket{u_{c'\vec k'}|\nabla_{\!\vec k'}u_{v'\vec k'}}^{*}
\Big].
\label{eq:interaction_term_before_sigma}
\end{align}

This is precisely the matrix element of the Hartree-Fock self-energy evaluated on the density-matrix variation \(\nabla_{\!\vec k} P_{\vec k}\). Indeed, for a general variation \(\delta P\),
\begin{equation}\label{eq:Sigma_HF_deltaP}
\Sigma^{\mathrm{HF}}_{mi}[\delta P]
=
\sum_{\alpha\beta}
\braket{m\alpha|\bar V|i\beta}\,
(\delta P)_{\beta\alpha}.
\end{equation}

Taking \(\delta P = \partial_\mu P\), we first evaluate the matrix elements of \(\partial_\mu P\). Since
\begin{equation}
P_{\vec k}=\sum_v \ketbra{u_{v\vec k}}{u_{v\vec k}}
\end{equation}
is a rank-\(N_{\mathrm{occ}}\) projector satisfying \(P_{\vec k}^2=P_{\vec k}\), its derivative is purely off-diagonal between occupied and empty subspaces:
\begin{equation}\label{eq:dP_offdiag}
\partial_\mu P_{\vec k}
=
\sum_{cv}
\left[
A^\mu_{cv}(\vec k)\ketbra{u_{c\vec k}}{u_{v\vec k}}
+
\big(A^\mu_{cv}(\vec k)\big)^*
\ketbra{u_{v\vec k}}{u_{c\vec k}}
\right],
\end{equation}
where
\begin{equation}\label{eq:A_definition}
A^\mu_{cv}(\vec k)
\equiv
\matrixel{u_{c\vec k}}{\partial_\mu P_{\vec k}}{u_{v\vec k}}
=
\braket{u_{c\vec k}|\partial_\mu u_{v\vec k}}.
\end{equation}

Substituting \(m=c\vec k\) and \(i=v\vec k\) into Eq.~\eqref{eq:Sigma_HF_deltaP}, we obtain
\begin{align}
&\Sigma^{\mathrm{HF}}_{c\vec k,\, v\vec k}[\partial_\mu P]
=
\sum_{\alpha\beta}
\braket{c\vec k,\alpha|\bar V|v\vec k,\beta}
(\partial_\mu P)_{\beta\alpha}
\nonumber\\
&=
\sum_{c'v'\vec k'}
\Big[
\braket{c\vec k,v'\vec k'|\bar V|v\vec k,c'\vec k'}
(\partial_\mu P)_{c'\vec k',\,v'\vec k'}
\nonumber\\
&\hspace{1cm}
+
\braket{c\vec k,c'\vec k'|\bar V|v\vec k,v'\vec k'}
(\partial_\mu P)_{v'\vec k',\,c'\vec k'}
\Big]
\nonumber\\
&=
\sum_{c'v'\vec k'}
\Big[
\braket{c\vec k,v'\vec k'|\bar V|v\vec k,c'\vec k'}
\braket{u_{c'\vec k'}|\partial_\mu u_{v'\vec k'}}
\nonumber\\
&\hspace{1cm}
+
\braket{c\vec k,c'\vec k'|\bar V|v\vec k,v'\vec k'}
\braket{u_{c'\vec k'}|\partial_\mu u_{v'\vec k'}}^{*}
\Big].
\label{eq:SigmaHF_dP_result}
\end{align}

 Eq.~\eqref{eq:SigmaHF_dP_result} is the same as  $1/\hbar$ multiply Eq.~\eqref{eq:interaction_term_before_sigma}, so the right hand side of Eq.~\eqref{eq:Gamma_q0_appendix} becomes
\begin{align}
&\vec v_{c\mathbf k,\, v\mathbf k}
+
\frac{1}{\hbar}
\matrixel{u_{c\mathbf k}}{\nabla_{\!\mathbf k}\Sigma^{\mathrm{HF}}[P]}{u_{v\mathbf k}}\\
=&
\frac{1}{\hbar}
\matrixel{u_{c\mathbf k}}{\nabla_{\!\mathbf k}H^0(\mathbf k)}{u_{v\mathbf k}}
+
\frac{1}{\hbar}
\matrixel{u_{c\mathbf k}}{\nabla_{\!\mathbf k}\Sigma^{\mathrm{HF}}[P]}{u_{v\mathbf k}}\\
=&
\frac{1}{\hbar}
\matrixel{u_{c\mathbf k}}{\nabla_{\!\mathbf k}H^{\mathrm{MF}}(\mathbf k)}{u_{v\mathbf k}}.
\end{align}
This shows that Eq.~\eqref{eq:Gamma_exact_solution} is a solution of the self-consistent vertex equation.

%Physically, $\nabla_{\vec k}P_{\vec k}$ describes the infinitesimal \emph{tilt} of the occupied subspace toward the unoccupied subspace as $\vec k$ is varied.  Because of the projector constraint, only particle-hole matrix elements contribute. The extent of this rotation is larger when the energy separation between occupied and unoccupied states is small.

 \section{Orbital Magnetization Formula}
We now show our formula for the orbital magnetization can be written as the so-called modern theory of orbital magnetization \cite{JShi_OM_2007, Thonhauser_OM_2005, Xiao_OM_2005, Ceresoli_OM_2006},  
\begin{align}
    M_z &= \frac{ie}{\hbar} \int\frac{d^2k}{4\pi^2} \sum_{c,v} 
    \frac{\xi_{\bk c} + \xi_{\bk v}}{2}\,
    \epsilon_{\alpha\beta}
    \braket{u_{\bk v}|\partial_\alpha u_{\bk c}}
    \braket{u_{\bk c}|\partial_\beta u_{\bk v}} \nonumber \\
    &= -\frac{ie}{2\hbar} \int\frac{d^2k}{4\pi^2} \sum_{c,v} \xi_{\bk c}\,
    \epsilon_{\alpha\beta}
    \braket{\partial_\alpha u_{\bk v}|u_{\bk c}}
    \braket{u_{\bk c}|\partial_\beta u_{\bk v}} \nonumber \\
    &\quad
    -\frac{ie}{2\hbar} \int\frac{d^2k}{4\pi^2} \sum_{c,v} \xi_{\bk v}\,
    \epsilon_{\alpha\beta}
    \braket{\partial_\alpha u_{\bk v}|u_{\bk c}}
    \braket{u_{\bk c}|\partial_\beta u_{\bk v}} .
\end{align}

To simplify the first line, we use,
\begin{align}
   \sum_{c} \xi_{\bk c} \ket{u_{\bk c}}\bra{u_{\bk c}}
   = H_{\bk} - \mu - \sum_{v'} \xi_{\bk v'} 
   \ket{u_{\bk v'}}\bra{u_{\bk v'}} ,
\end{align}
which gives
\begin{align}
   &\sum_{c,v} \xi_{\bk c}\,\epsilon_{\alpha\beta}
   \braket{\partial_\alpha u_{\bk v}|u_{\bk c}}
   \braket{u_{\bk c}|\partial_\beta u_{\bk v}}\\
   &= \sum_{v} \epsilon_{\alpha\beta}
   \bra{\partial_\alpha u_{\bk v}}
   \!\left(
   H_{\bk} - \mu - \sum_{v'} \xi_{\bk v'}
   \ket{u_{\bk v'}}\bra{u_{\bk v'}}
   \right)
   \!\ket{\partial_\beta u_{\bk v}} .
\end{align}

Similarly, for the second line we use
\begin{align}
   \sum_{c} \ket{u_{\bk c}}\bra{u_{\bk c}}
   = \mathbb{1} - \sum_{v'} \ket{u_{\bk v'}}\bra{u_{\bk v'}} ,
\end{align}
which gives,
\begin{align}
   &\sum_{c,v} \xi_{\bk v}\,\epsilon_{\alpha\beta}
   \braket{\partial_\alpha u_{\bk v}|u_{\bk c}}
   \braket{u_{\bk c}|\partial_\beta u_{\bk v}}\\
  & = \sum_{v} \xi_{\bk v}\,\epsilon_{\alpha\beta}
   \bra{\partial_\alpha u_{\bk v}}
   \!\left(
   \mathbb{1} - \sum_{v'} \ket{u_{\bk v'}}\bra{u_{\bk v'}}
   \right)
   \!\ket{\partial_\beta u_{\bk v}} .
\end{align}

Adding these two expressions yields a term
\begin{equation}
    \sum_{v,v'} (\xi_{\bk v} + \xi_{\bk v'})\,
    \epsilon_{\alpha\beta}\,
    \braket{\partial_\alpha u_{\bk v}|u_{\bk v'}}\,
    \braket{u_{\bk v'}|\partial_\beta u_{\bk v}} = 0 ,
\end{equation}
because \((\xi_{\bk v} + \xi_{\bk v'})\) is symmetric in \(v,v'\),
while 
\(\epsilon_{\alpha\beta}\braket{\partial_\alpha u_{\bk v}|u_{\bk v'}}\braket{u_{\bk v'}|\partial_\beta u_{\bk v}}\)
is antisymmetric. Therefore, we arrive at the gauge-invariant expression for the orbital magnetization,
\begin{align}
    M_z = 
    \frac{e}{2\hbar} \int\frac{d^2k}{4\pi^2} 
    \sum_{v}
    \epsilon_{\alpha\beta}
    \Im \!
    \left[
    \bra{\partial_\alpha u_{\bk v}}
    \left( H_{\bk} + \varepsilon_{\bk v} - 2\mu \right)
    \ket{\partial_\beta u_{\bk v}}
    \right] .
\end{align}

\end{document}